\documentclass[aps,prb,twocolumn,notitlepage,showpacs,superscriptaddress,10pt]{revtex4-2}%
\usepackage{graphicx}
\usepackage{amsmath}
\usepackage{amssymb}
\usepackage{color}
\usepackage{amsfonts}%
\usepackage[caption=false]{subfig}

\usepackage{color}
\usepackage[colorlinks,bookmarks=false,citecolor=darkblue,linkcolor=red,urlcolor=blue]{hyperref}

\definecolor{darkred}{rgb}{0.7,0.0,0.0}

\definecolor{darkblue}{rgb}{0,0.02,0.45}

\definecolor{darkgreen}{rgb}{0.02,0.45,0.0}

\definecolor{violet}{rgb}{0.8,0.2,0.6}

\providecommand{\U}[1]{\protect\rule{.1in}{.1in}}
\newcommand{\ALIO}{Ag$_3$LiIr$_2$O$_6$}

\newcommand{\AIO}{A$_2$IrO$_3$}
\newcommand{\LIO}{$\alpha$-Li$_2$IrO$_3$}

\newcommand{\CIO}{Cu$_2$IrO$_3$}

\begin{document}
\title{Origin of the insulating state in the honeycomb iridate {\CIO} }

%\title{Origin of the insulating state in the Kitaev candidate {\CIO} }
%\title{Insulating behavior and binary magnetic ions in the intercalated Kitaev candidate \CIO~from first principles}
%\title{Origin of the insulating behavior in the Kitaev candidate \CIO~from first principles}
%
%\title{Origin of the insulating state in the Kitaev candidate {\CIO} }
%\title{Iridium charge order in the Kitaev candidate \CIO}
%insulating states for Cu2IrO3 from first principles
\author{Ying Li}\thanks{yingli1227@xjtu.edu.cn}
\affiliation{Department of Applied Physics and MOE Key Laboratory for Nonequilibrium Synthesis and Modulation of Condensed Matter, School of Physics, Xi'an Jiaotong University, Xi'an 710049, China}
\author{Roger~D. Johnson} 
\affiliation{Department of Physics and Astronomy, University College London, Gower Street, London WC1E 6BT, United Kingdom}
\author{Yogesh Singh} 
\affiliation{Indian Institute of Science Education and Research, Mohali, Sector 81, SAS Nagar, Manauli 140306, India}
\author{Radu Coldea}
\affiliation{Clarendon Laboratory, University of Oxford, Physics Department, Oxford OX1 3PU, United Kingdom}
\author{Roser Valent{\'\i}}\thanks{valenti@itp.uni-frankfurt.de}
\affiliation{Institut f\"ur Theoretische Physik, Goethe-Universit\"at Frankfurt,
Max-von-Laue-Strasse 1, 60438 Frankfurt am Main, Germany}
\date{\today}

\begin{abstract}
Through a combination of crystal symmetry analysis and density functional theory calculations we unveil a possible microscopic origin of the unexpected insulating behavior reported in the
honeycomb Kitaev material {\CIO}. Our study suggests that this material hosts an instability towards charge ordering of the Ir ions, with alternating magnetic Ir$^{4+}$ and non-magnetic Ir$^{3+}$ ions arranged on the honeycomb lattice. In this case, the next-nearest-neighbor interactions that couple magnetic Ir$^{4+}$ ions form an enlarged triangular lattice, instead of the expected honeycomb lattice. The magnetic Cu$^{2+}$ ions located at the centre of the iridium honeycomb voids also form a triangular lattice, and additionally contribute to the magnetization of the system. Together, the interpenetrated Ir$^{4+}$ and Cu$^{2+}$ triangular lattices present a novel type of honeycomb Kitaev lattice composed of two types of magnetic ions. 
\end{abstract}

\maketitle

\par

\section{Introduction}
The iridate family {\AIO} (A = Na, Li) has been considered as a prime candidate 
to realize the long-sought Kitaev $\mathbb{Z}_2$ spin liquid in a honeycomb lattice with nearest neighbor bond-dependent Ising interactions~\cite{Kitaev2006,Jackeli2009,Chaloupka2013, Witczak-Krempa2014,Rau16,Schaffer2016,WinterReview,Trebst2017,Cao2018,Singh2010, Choi2012, Singh2012, Singh2012,Gretarsson2013,modic2014,Freund2016,Williams2016,Biffin2014,Biffin2014gamma,Takayama2015,Li2017}. However, the systems show long-range magnetic order due to the presence of further non-Kitaev interactions~\cite{Katukuri2014,Rau2014,Winter2016,Natalia2018,Natalia2018b}. Attempts to modulate the magnetic couplings have been pursued by
intercalation of H atoms~\cite{o2012production,Bette2017,Kitagawa2018} or Ag atoms~\cite{Bahram2019} in {\LIO}. In the former, theoretical studies~\cite{Li2018,Yadav-Hozoi2018h3liir2o6} indicated that H positions strongly affect the magnetic interactions~\cite{Li2018} resulting in magnetic models with bond disorder.
Such models in the presence of vacancies have been shown to reproduce the experimentally observed low-energy spectrum in H$_3$LiIr$_2$O$_6$ ~\cite{Knolle2019, Kao2021}. The iridate {\ALIO} was initially proposed to be closer to the Kitaev limit compared to \LIO, however by
improving  the sample quality, the system shows long range incommensurate antiferromagnetic (AFM) order~\cite{Bahrami2021, Wang2020}. In theoretical studies it was found that
the Ir-O hybridization in {\ALIO} is moderate and a localized relativistic $j_{\rm eff} = 1/2$ magnetic model
with Kitaev and non-Kitaev exchange contributions is still valid for the description of the system~\cite{li2022role}, albeit its properties~\cite{Torre2021} may be affected
by the presence of Ag vacancies~\cite{li2022role,Torre2021}. A relatively new intercalated honeycomb system {\CIO} has also been synthesized~\cite{Abramchuk2017} consisting of %nearly perfect 
Ir honeycomb layers, with Cu atoms situated both at the center of the honeycomb voids, and in between the Ir layers as shown Fig.~\ref{fig:structure} (a).
 
A crystal structure with $C2/c$ space group symmetry was first proposed for this system, but it was later shown that three structures with the same qualitative atomic connectivity but different space group symmetries, $C2/c$, $C2/m$, and $P2_1/c$, could not be uniquely distinguished using the published powder X-ray diffraction data~\cite{Fabbris2021}. Susceptibility and specific heat measurements show that the system remains magnetically disordered until 2.7 K, at which point short-range order develops~\cite{Abramchuk2017}. ${\mu}$SR measurements detect a two-component depolarization with slow- and fast-relaxation rates, attributed to co-existence of static and dynamic magnetism, with further evidence that both Ir$^{4+}$ and Cu$^{2+}$ magnetic moments exist~\cite{Kenney2019}. Taken together, these magnetic properties were suggested to arise from significant levels of chemical disorder~\cite{Choi2019}. However, signatures characteristic of a Kitaev quantum spin liquid have also been reported based on nuclear quadrupole resonance and Raman scattering measurements~\cite{Takahashi2019, Pal2021}. Furthermore, {\CIO} has been reported to show a complex set of structural phase transitions under hydrostatic pressure~\cite{Fabbris2021,Pal2022} and an insulator-to-metal transition~\cite{Jin2022}. More recently, Ref.~\cite{Haraguchi2024} reported a $C2/m$ crystal structure with less antisite disorder between Cu and Ir ions in the honeycomb layers. In this publication magnetic susceptibility measurements revealed a weak ferromagnetic-like anomaly with hysteresis at a magnetic transition temperature of ~70 K~\cite{Haraguchi2024}. The crystal structures and bulk properties summarised above are consistent with linearly-bonded, non-magnetic Cu$^{+}$ (3$d^{10}$) ions located between honeycomb layers, and octahedrally coordinated, magnetic Cu$^{2+}$ (3$d^9$) ions in the honeycomb voids. Charge is then balanced by a fractional oxidation state of Ir$^{3.5+}$ (5$d^{5.5}$), which would be expected to render the system metallic. It is hence an unresolved challenge to reconcile this expected metallic state with the experimentally observed insulating behavior at ambient pressure~\cite{Abramchuk2017}.

In this paper, we use symmetry considerations and density functional theory (DFT) to investigate the structural and electronic properties of {\CIO}. We take as a starting point a $C2/m$ crystal structure, which is consistent with several other monoclinic layered honeycomb materials such as Na$_2$IrO$_3$ \cite{Choi2012} and $\alpha$-Li$_2$IrO$_3$ \cite{o2012production}. Our structural relaxation and energy minimization results show that the insulating state in {\CIO} may result from iridium charge ordering into magnetic Ir$^{4+}$ (5$d^{5}$) and non-magnetic Ir$^{3+}$ (5$d^{6}$). The ideal magnetic ground state is therefore based on a honeycomb Kitaev lattice, composed not of nearest-neighbour Ir$^{4+}$ ions, but of alternating Ir$^{4+}$ and Cu$^{2+}$ ions with non-magnetic Ir$^{3+}$ in the honeycomb voids.

\section{Crystal Structure}\label{sec:struct}
\begin{figure}[tpb]
\center
\includegraphics[angle=0,width=\linewidth]{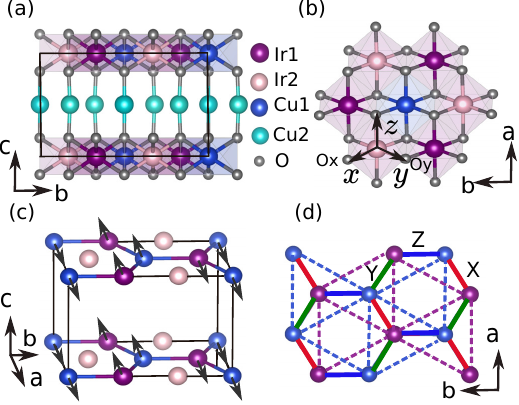}
\caption{Proposed charge-ordered crystal structure of {\CIO} ($C_2$ symmetry) projected into (a) the $bc$- and (b) the $ab$- planes. The two nonequivalent Ir ions are labeled as Ir1 and Ir2. Cu1 and Cu2 indicate the copper ions in the honeycomb layer and between the layers, respectively. Ir1, Ir2 and Cu1 label single crystallographic sublattices, while Cu2 labels ions located on 3 symmetry-independent sublattices indicated as Cu2-1, Cu2-2, and Cu2-3 in Table~\ref{c2} in Appendix A. The average distances Ir1-O and Ir2-O are 2.04 \AA~and 2.07 \AA, respectively. $x$, $y$, $z$ are Cartesian axes used to describe the $d$-orbitals of all Ir and Cu ions; $x$ and $y$ are defined with respect to the Ir2 octahedra, with the ${\bf{\hat{x}}}+{\bf{\hat{y}}}$ diagonal bisecting the Ox-Ir-Oy bond angle and the ${\bf{\hat{x}}}-{\bf{\hat{y}}}$ diagonal along the 2-fold $b$-axis. $z$ then completes the right-handed orthonormal set, perpendicular to the $xy$ plane. (c) The `zig-zag' antiferromagnetic configuration used in the GGA+SO+$U$ calculations. (d) Honeycomb lattice containing magnetic Cu1 (blue) and Ir1 (purple) ions. The Cu1-Ir1 distances are 3.11~\AA, and triangular lattices of magnetic Ir ions (Ir1) and Cu ions (Cu1) are shown by dashed lines. The Ir-Ir and Cu-Cu distances are 5.4~\AA.}
\label{fig:structure}
\end{figure}

The crystal structure of {\CIO} is composed of honeycomb layers formed from edge sharing IrO$_6$ octahedra, with Cu atoms occupying both the honeycomb voids (Cu1) and the space between the layers (Cu2); see Fig.~\ref{fig:structure}. We use as starting point a crystal structure with $C2/m$ symmetry derived from the $C2/c$ structure proposed in Ref.~\cite{Abramchuk2017}. This model was further refined by performing a structural relaxation (see details below) in which the atomic fractional coordinates were allowed to vary, but the unit cell parameters were kept fixed to the experimental values \cite{Abramchuk2017} (the $C2/m$ unit cell (unprimed) is related to the $C2/c$ cell (primed) by the transformation $\mathbf{a} = -\mathbf{a}'$, $\mathbf{b} = -\mathbf{b}'$, $\mathbf{c} = (\mathbf{c}' + \mathbf{a}')/2$). The resulting parameters are given on the right side of Table~\ref{c2} in Appendix A, and were found to be in good agreement with those of the $C2/m$ structure recently published in Ref.~\cite{Haraguchi2024}. The octahedral coordination of the Cu1 sites is usually compatible with a Cu$^{2+}$ valence state, while the linear bonding of the Cu2 sites is typical of a Cu$^{+}$ valence. This Cu charge configuration was confirmed by bond valence sum calculations performed on our relaxed $C2/m$ structure. Charge neutrality then implies an average iridium oxidation state of $+3.5$, i.e. a nominal composition Cu$^{2+}_{0.5}$Cu$^+_{1.5}$Ir$^{3.5+}$O$_3^{2-}$.    

In $C2/m$ symmetry, the iridium ions are located on a single sublattice of symmetry-equivalent sites, so a fractional oxidation state of $+3.5$ for all iridium sites would be expected to lead to a metallic behaviour. However, this fractional oxidation state also introduces an instability towards charge disproportionation, which would open an energy gap at the Fermi level giving rise to an insulator. Long-range charge order comprised of alternating nearest-neighbour Ir$^{3+}$ and Ir$^{4+}$ ions breaks both mirror and inversion symmetry. There is just one maximal subgroup of $C2/m$ compatible with this broken symmetry; $C2$. While other, lower symmetry subgroups may be realised, we limit our discussion to the highest symmetry case as is typical in the study of symmetry-breaking order. Any charge order will be accompanied by an ordered pattern of atomic displacements allowed within $C2$ symmetry. In particular, one would expect to find small oxygen displacements that expand and contract the Ir$^{3+}$ and Ir$^{4+}$ octahedra, respectively. 

To test the hypothesis of a charge ordered ground state we performed a structural relaxation within a model with $C2$ symmetry using the Vienna {\it ab initio} simulation package (VASP)~\cite{Kresse1996,Hafner2008}. We considered relativistic effects as well as contributions of the Coulomb repulsion~\cite{Dudarev1998} [$U_{\rm eff}$ = $U - J_H$ = 2.4 eV for Ir ($U^{\rm Ir}_{\rm eff}$) following calculations for Na$_2$IrO$_3$~\cite{Li2015} and 8 eV for Cu ($U^{\rm Cu}_{\rm eff}$) following calculations for ZnCu$_3$(OH)$_6$Cl$_2$~\cite{Pustogow2017}] within GGA+SO+$U$. We adopted a cutoff energy of 520 eV and a Monkhorst-pack 4 $\times$ 2 $\times$ 4 $k$-points mesh. The structure was initialised with a nominal, symmetry breaking distortion where the Ir1-O bond lengths were shortened on average $(\bar{r}_s)$, and the Ir2-O bond lengths were lengthened on average $(\bar{r}_l)$. The lattice parameters were fixed and the fractional coordinates of all ions were allowed to vary. The resultant structural parameters are given in the left side of Table \ref{c2} in Appendix A (see also Fig. \ref{fig:structure}). The differences in relaxed Cu and Ir positions between the $C2/m$ reference structure and the $C2$ structure were negligible, while the average Ir-O bond lengths were found to have a ratio of $\bar{r}_s/\bar{r}_l$ $\sim$ 0.98. The respective small shifts in oxygen positions are likely within the uncertainty of structural refinements against X-ray diffraction data, especially given the presence of stacking faults typically found in monoclinic layered honeycomb materials. However, our DFT calculations showed that the respective electronic properties are strongly affected, whereby the difference between the two iridium-oxygen coordinations induces a full charge order with nearest neighbour Ir$^{4+}$ (Ir1) and Ir$^{3+}$ (Ir2) ions. The total energy difference between $C2/m$ and $C2$ structural relaxations was only 4~meV per formula unit. Despite being close to the accuracy limit of our DFT calculations, this energy difference was found to be significant. Remarkably, such a small change in energy corresponded to complete charge order of the iridium ions, suggesting that slight perturbations may ultimately prevent long-range charge order in the real material. In this case the ground state could resemble a disordered, glassy array of Ir$^{4+}$ and Ir$^{3+}$ ions with their respective, disordered oxygen displacements. We note that the average symmetry of the disordered ground state is $C2/m$, and therefore may be easily hidden in any diffraction experiment. For completeness, we also tested charge ordered structures based on the other two experimentally suggested symmetries ($P2_1/c$ and $C2/c$), and found robust charge order and insulating behavior in these symmetries.

\section{Electronic properties}
\begin{figure}
\center
\includegraphics[angle=0,width=\linewidth]{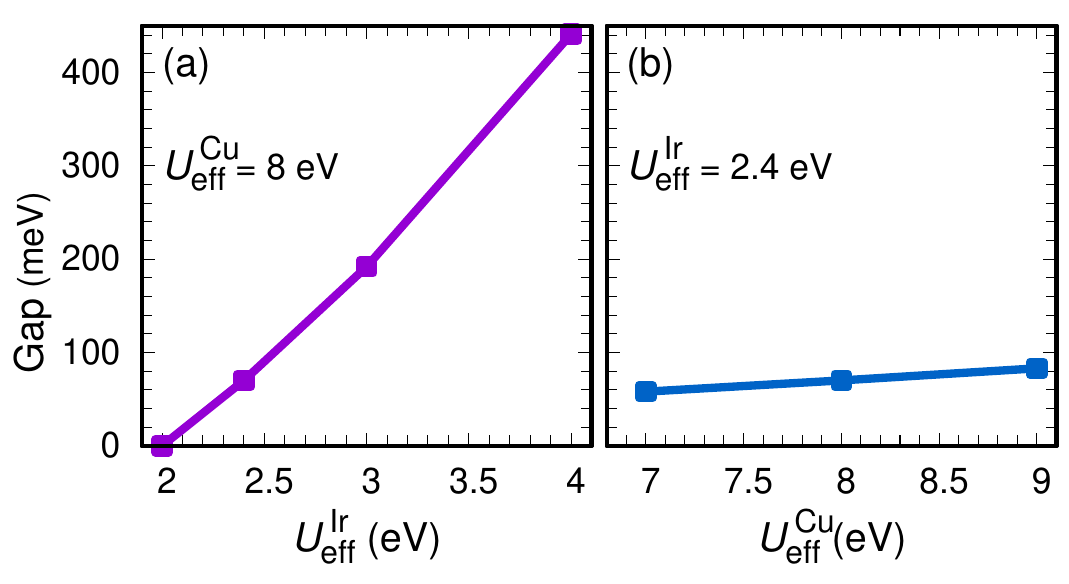}
\caption{Insulating gap in {\CIO} as a function of (a) $U^{\rm Ir}_{\rm eff}$ ($U^{\rm Cu}_{\rm eff}$ = 8 eV) and (b) $U^{\rm Cu}_{\rm eff}$ ($U^{\rm Ir}_{\rm eff}$ = 2.4 eV) using GGA+SO+$U$ with an antiferromagnetic configuration as shown in Fig~\ref{fig:structure} (c).
 $U^{\rm Ir}_{\rm eff}$ and $U^{\rm Cu}_{\rm eff}$ are the onsite Coulomb repulsion in Ir and Cu, respectively.}
\label{fig:gapu}
\end{figure}

The electronic properties were obtained from full-potential linearized augmented plane-wave (LAPW) calculations~\cite{Wien2k}. We chose the basis-size controlling parameter RK$_{\rm max}$ = 7 and a mesh of 500 {\bf k} points in the first Brillouin zone (FBZ) of the primitive unit cell. The density of states (DOS) were computed with 1000 {\bf k} points in the full Brillouin zone. An analysis of the insulating gap as a function of $U_{\rm eff} = U - J_{\rm H}$ was performed, having established that a `zig-zag' antiferromagnetic configuration (see Fig~\ref{fig:structure}c) gave the lowest total energy compared to three other commensurate magnetic configurations (ferromagnetic, Néel, and stripy). As shown in Fig.~\ref{fig:gapu}, increasing $U^{\rm Ir}_{\rm eff}$ from 2 eV to 4 eV, the charge gap increases sharply from 0 to 0.44 eV. With increasing $U^{\rm Cu}_{\rm eff}$ from 7 eV to 9 eV, the gap increases slightly from 0.058 eV to 0.083 eV. The $U_{\rm eff}$ values for iridium atoms clearly affect the energy gap much more than those of copper. 

\begin{figure}
\center
\includegraphics[angle=0,width=\linewidth]{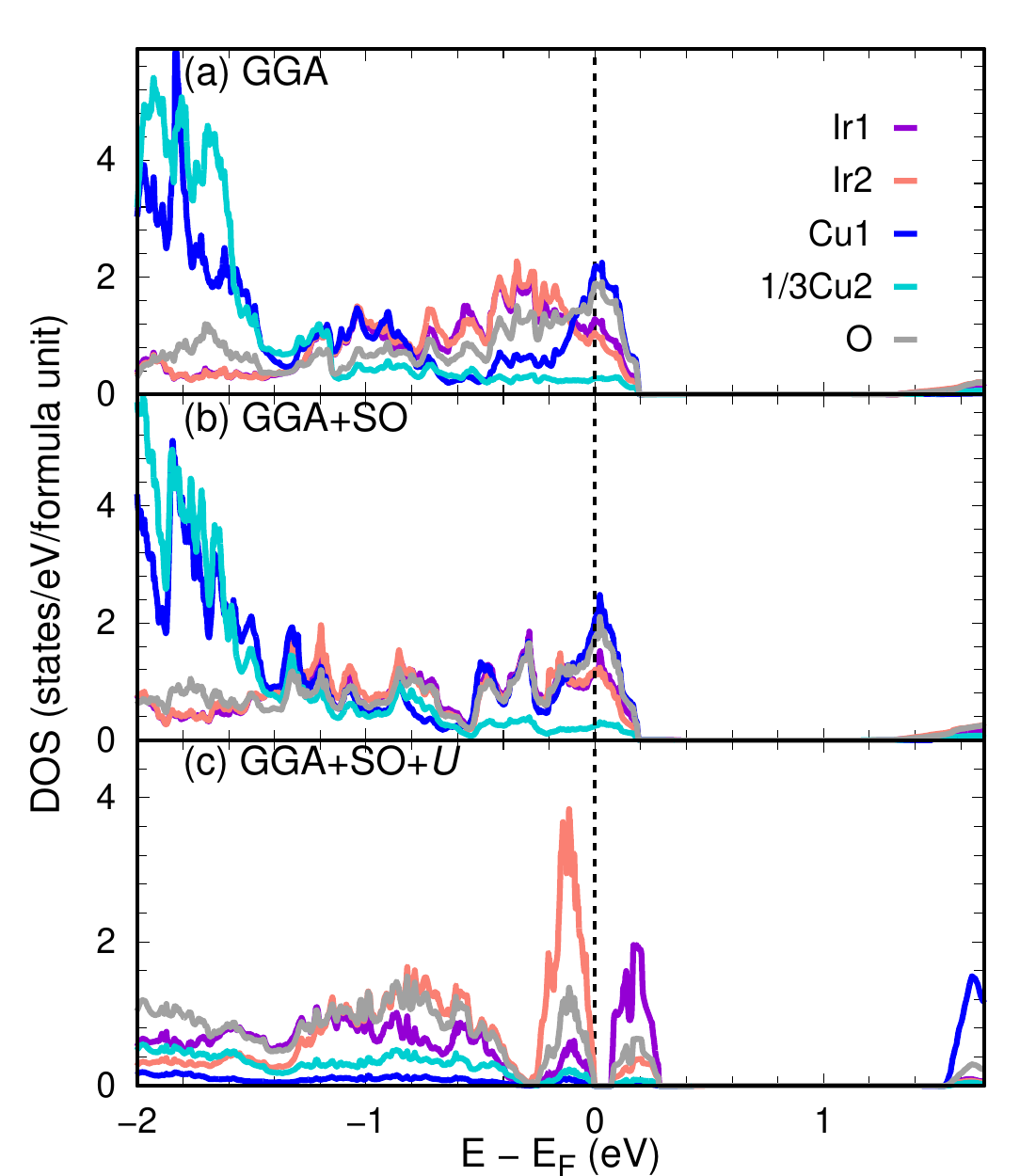}
\caption{Density of states obtained from (a) nonmagnetic GGA, (b) nonmagnetic GGA+SO and (c) GGA+SO+$U$ with the magnetic configuration displayed in Fig.~\ref{fig:structure} (c). Note that the Cu2 DOS has been scaled by a factor 1/3.}
\label{fig:pdos}
\end{figure}

In Fig~\ref{fig:pdos} we show the partial density of states (DOS) for the relxaed $C2$ structure obtained within nonmagnetic GGA and GGA+SO approximations, as well as GGA+SO+$U$ assuming the zig-zag antiferromagnetic configuration. In both the nonrelativistic GGA (Fig~\ref{fig:pdos}a) and relativistic GGA+SO (Fig~\ref{fig:pdos}b) partial DOS, Cu2 $d$ states are almost fully occupied below the Fermi level, while Cu1 $d$ states are partially occupied around the Fermi level. This result is consistent with Cu$^{+}$ at the Cu2 site and Cu$^{2+}$ at the Cu1 site, as expected. The partial DOS of Ir1 and Ir2 differ in magnitude around the Fermi level, but have approximately the same form. As shown in Fig~\ref{fig:pdos}c, including a Coulomb repulsion for Ir and Cu (GGA+SO+$U$) induces a charge order of Ir1 and Ir2, and opens a gap of about 70 meV (with $U_{\rm eff} = U - J_H$ = 2.4 eV~\cite{Li2015}) for Ir and 8 eV for Cu. We note that GGA+$U$ calculations without spin-orbit coupling, not shown here, converged to a metallic ferrimagnetic state where Ir1 has bigger moment than Ir2. Therefore, spin-orbit coupling, Coulomb repulsion, and magnetic moments on Ir are all important factors to stabilize the charge ordered state and the respective insulating behavior.

\begin{figure}
\center
\includegraphics[angle=0,width=\linewidth]{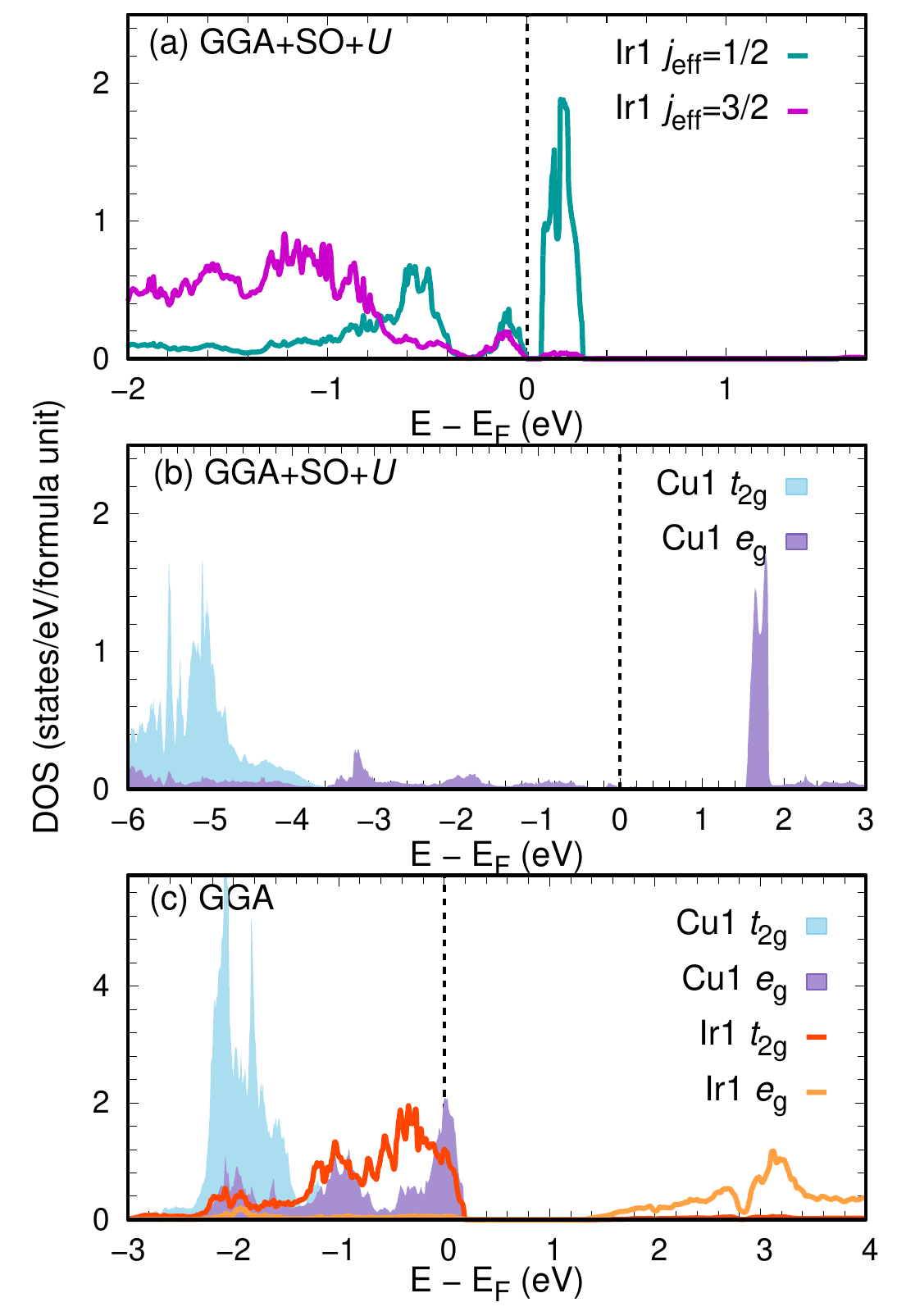}
\caption{The GGA+SO+$U$ density of states projected onto (a) the relativistic $j_{\rm eff}$ basis of Ir1, and (b) the $t_{2g}$ and $e_g$ orbitals of Cu1. (c) Nonrelativistic GGA density of states projected onto $t_{2g}$ and $e_g$ orbitals for both Cu1 and Ir1.}
\label{fig:pdosircu}
\end{figure}

To further clarify the microscopic nature of the insulating state, we re-evaluate the previously calculated GGA+SO+$U$ Ir1 density of states projected onto a relativistic $j_{\rm eff}$ basis (Fig. \ref{fig:pdos}a). The Cu1 density of states was also projected onto a $t_{2g}$ and $e_g$ basis (Fig.~\ref{fig:pdosircu}b). These data show that Ir1 has one hole in the $j_{\rm eff} = 1/2$ state and Cu1 has one hole in the $e_g$ orbital, both of which therefore contribute to the magnetism of the compound.

\section{Magnetic properties\label{sec::mag}}
We derived the magnetic exchange parameters using the projED method~\cite{Winter2016, Riedl2019}, which consists of the following two steps. First, {\it ab-initio} hopping parameters between the Ir and Cu ions are extracted with projective Wannier functions \cite{Eschrig2009} applied to non relativistic FPLO~\cite{Koepernik1999} calculations on a $10 \times 10 \times 10$ $\mathbf{k}$ mesh. This allows us to construct an effective electronic model, $\mathcal{H}_{\rm tot} = \mathcal{H}_{\rm hop}+  \mathcal{H}_{\rm SO} + \mathcal{H}_{\rm U}$, that includes the above calculated kinetic hopping term $\mathcal{H}_{\rm hop}$, plus the spin-orbit coupling $\mathcal{H}_{\rm SO}$ and Coulomb interaction $\mathcal{H}_{U}$ contributions. The spin-orbit coupling $\lambda$ is set to 0.4 eV for Ir, and the Coulomb interactions were set to $U^{\rm Ir}$ = 1.7 eV, $J^{\rm Ir}_{\rm H}$ = 0.3 eV, $U^{\rm Cu}$ = 8 eV, and $J^{\rm Cu}_{\rm H}$ = 1 eV. We note that the Coulomb interaction value for Ir~\cite{Winter2016} considered in this effective Hubbard model is smaller than the one used in the GGA+U+SO calculations within the LAPW implementation, as we have explained in previous calculations~\cite{Kaib2022}. In a next step, the effective spin Hamiltonian $\mathcal{H}_{\rm eff}$ is extracted from the electronic model via exact diagonalization (ED) of finite clusters. The projection of the resulting energy spectrum onto the low-energy subspace is then obtained by adopting a pseudo-spin operator representation in the $j_{\rm eff}$ basis for Ir, and an $e_g$ basis for Cu, with the projection operator defined as $\mathbb{P}$: $\mathcal{H}_{\rm eff} = \mathbb{P} \mathcal{H}_{\rm tot}\mathbb{P} = \sum_{i j } \mathbf{S}_i\, \mathbb{J}_{ij}\, \mathbf{S}_j$. 

Considering only the magnetic ions Ir1 and Cu1, the nearest neighbour interactions between Ir1 and Cu1 span bonds labeled X, Y, and Z (see Fig.~\ref{fig:structure}d). In the $C2$ space group symmetry, X and Y bonds are related by 2-fold rotation about the crystallographic \textit{b}-axis, and the Z bonds are independent. Strong hybridization of Ir and Cu was found in our calculations, which is consistent with the enhanced delocalization of $5d$ $j_{\rm eff}$ = 1/2 orbitals observed experimentally by resonant inelastic X-ray scattering~\cite{Fabbris2024}. The next-nearest-neighbour interactions are Ir1-Ir1 and Cu1-Cu1, which form two interpenetrating triangular lattices. The respective interaction distances are around 5.5~\AA, much larger than the Ir1-Cu1 nearest-neighbour distance. Hence, in the following minimal model we only consider the hopping parameters between nearest neighbors Ir1 and Cu1.

The obtained crystal fields for Ir1 and Cu1 ions are shown in Table~\ref{tab:cef}. For the onsite orbital levels of Ir1 ions, $e_g$ are empty above the Fermi level lying at 2.7 eV while $t_{2g}$ are occupied at around -0.5 eV. For Cu1, the $t_{2g}$ orbitals are fully occupied and located at around -1.9 eV while the $e_g$ orbitals are found around -0.5 eV. These results are consistent with the GGA density of states presented in Fig.~\ref{fig:pdosircu}c. Therefore the orbitals related to the electronic and magnetic properties close to the Fermi level are $t_{2g}$ orbitals of Ir1 and $e_g$ orbitals of Cu1. 

\begin{table}[]
    \centering\def\arraystretch{1.1}
       \caption{Matrix elements of the crystal field Hamiltonian for Ir1 and Cu1 in units of meV, given in the $t_{2g}$, $e_g$ basis defined with respect to the Cartesian axes $xyz$ defined in Fig.~\ref{fig:structure}b). Values in bold indicate the dominant terms. The form of the Cu1 crystal field is consistent with its 2-fold site symmetry along the ${\bf{\hat{x}}}-{\bf{\hat{y}}}$ direction. The Ir1 site has the same symmetry, and the additional zeros in the matrix Hamiltonian are likely due to the large energy gap between $t_{2g}$ and $e_g$ states.}
    \begin{ruledtabular}
    \begin{tabular}{l|r|rrrrr}
   & & $d_{xy}$ & $d_{xz}$ & $d_{yz}$ & $d_{z^2}$ &  $d_{x^2\text{-}y^2}$ \\
    \hline
   &$d_{xy}$ & \textbf{-542.4} & -7.8 &  -7.8   & 0 & 0 \\
   &$d_{xz}$ & -7.8 & \textbf{-548.7}  & -13.7  & 0 &  0\\   
  Ir1 &$d_{yz}$ & -7.8  & -13.7 &  \textbf{-548.7} & 0 & 0\\  
   &$d_{z^2}$ & 0 & 0 & 0 & \textbf{2665.7} & 0\\
   &$d_{x^2\text{-}y^2}$ & 0 &  0 & 0 & 0 & \textbf{2689.1}\\   
   \hline
  % On-site Cu & d_{xy} & d_{xz} & d_{yz} & d_{z^2} &  d_{x^2\text{-}y^2} \\
   &$d_{xy}$ &  \textbf{-1884.1} & -10.0 & -10.0 & 71.0 & 0 \\
   &$d_{xz}$  &  -10.0 & \textbf{-1885.6} & 7.2 & -116.4 &  0.1 \\   
   Cu1 &$d_{yz}$  & -10.0 & 7.2 & \textbf{-1885.6} & -116.4 & -0.1\\   
   &$d_{z^2}$  & 71.0 & -116.4 & -116.4 & \textbf{-526.7} & 0\\
   &$d_{x^2\text{-}y^2}$ & 0  &  0.1  &  -0.1 & 0  & \textbf{-543.3}\\
    \end{tabular}
    \end{ruledtabular}
    \label{tab:cef}
\end{table}

The calculated, bond-dependent nearest-neighbour hopping parameters of Ir1-Cu1 bonds are given Table~\ref{tab:hop}, again in the $t_{2g}$, $e_{g}$ orbital basis. For the Z-bond, hopping between Ir1 $d_{xy}$ and Cu1 $d_{z^2}$ was found to dominate, while for the X(Y)-bond, the significant hopping contributions were between Ir1 $d_{yz}$ ($d_{xz}$) and Cu1 $d_z^{2}$ and $d_{x^2-y^2}$ orbitals. 

\begin{table}[]
    \centering\def\arraystretch{1.1}
       \caption{Hopping parameters for Ir1-Cu1 nearest neighbour bonds, in units of meV, calculated for the X, Y, and Z bonds. Values in bold indicate the dominant terms. Note that the parameters found for X and Y bonds are consistent with the 2-fold rotational symmetry around ${\bf{\hat{x}}}-{\bf{\hat{y}}}$ that inter-relates these two bonds.}
    \begin{ruledtabular}
    \begin{tabular}{lr|rrrrr}
   & & \multicolumn{5}{c}{Cu1} \\
   & & $d_{xy}$ & $d_{xz}$ & $d_{yz}$ & $d_{z^2}$ &  $d_{x^2\text{-}y^2}$ \\
    \hline    
   %Ir-Cu(H) (Z-bond) & d_{xy} & d_{xz} & d_{yz} & d_{z^2} &  d_{x^2\text{-}y^2} \\ 
   Ir1 (Z)&$d_{xy}$ &13.1 & -9.0 &-9.0 &\textbf{231.1} & 0\\  
   &$d_{xz}$  & -0.6  & -11.8 &-1.7 &0.1 &-28.8\\   
   &$d_{yz}$ & -0.6  & -1.7  & -11.8 &  0.1 & 28.8\\   
   &$d_{z^2}$ & 0 & 0 & 0 & 0 & 0 \\
   &$d_{x^2\text{-}y^2}$ & 0 & 0 & 0 & 0 & 0 \\ 
   \hline
    Ir1 (Y)&$d_{xy}$ & -9.6  &-5.6 &-1.8  & 46.3  & 10.4 \\      
   &$d_{xz}$  &  -0.7  & 11.6  & 13.5  & \textbf{-109.1} & \textbf{-196.6} \\   
    &$d_{yz}$ & -1.9  & 2.8   & -9.8  & -21.9  & 32.0  \\
   &$d_{z^2}$ & 0 & 0 & 0 & 0 & 0 \\
   &$d_{x^2\text{-}y^2}$ & 0 & 0 & 0 & 0 & 0 \\  
   \hline
    Ir1 (X)&$d_{xy}$ & -9.6 & -1.8 & -5.6 & 46.3 & -10.4\\
   &$d_{xz}$  &  -1.9 & -9.8 & 2.8 & -21.9 & -32.0\\
   &$d_{yz}$ & -0.7 & 13.5 & 11.6 & \textbf{-109.1} & \textbf{196.6}\\   
   &$d_{z^2}$ & 0 & 0 & 0 & 0 & 0 \\
   &$d_{x^2\text{-}y^2}$ & 0 & 0 & 0 & 0 & 0 \\ 
    \end{tabular}
    \end{ruledtabular}
    \label{tab:hop}
\end{table}

The resulting effective Hamiltonian can be written as:
\begin{align}
\mathcal{H}_{\rm spin} = \sum_{\langle ij\rangle} \mathbf{S}_i \cdot \mathbf{J}_{ij} \cdot \mathbf{S}_j 
\label{eqn:spin_ham}
\end{align}
where $\langle ij\rangle$ denotes a sum over all pairs of nearest-neighbour sites. In Kitaev's original honeycomb model, the exchange parameters are bond-dependent. For example, the Z-bond interaction can be written as
\begin{align}
\mathbf{J}^Z=\left(\begin{array}{ccc} J + \xi& \Gamma & \Gamma^{\prime} + \zeta\\ 
\Gamma  & J - \xi & \Gamma^{\prime} - \zeta \\ 
\Gamma^{\prime} + \zeta & \Gamma^{\prime} - \zeta & J+K \end{array}\right).
\end{align}
The calculated exchange parameters for all three bonds are given in Table~\ref{tab:mag}.  
For the Z bond this magnetic model has a large negative Heisenberg $J$ and a positive Kitaev $K$ term; both of those terms change sign and reduce in magnitude for the X and Y bonds. A mean-field calculation using SpinW~\cite{Toth2015} finds that this Hamiltonian has a Néel-type magnetic order in the ground state, which differs from the zig-zag order found in our DFT total energy calculations. We suggest that this discrepancy may come from the presence of significant magnetic interactions beyond the minimal nearest-neighbour model in (\ref{eqn:spin_ham}), such as antiferromagnetic next-nearest neighbour couplings, which would favour zig-zag order. Inclusion of these terms in the model calculations require larger cluster diagonalization~\cite{Winter2016}, which, due to the complexity of the present system with two different magnetic ions, are left for future studies.

\begin{table}[t]
\caption {Calculated nearest-neighbor exchange parameters in units of meV.}
\centering\def\arraystretch{1.1}
\label{tab:mag}
\begin{ruledtabular}
\begin{tabular}{l|rrrrrrr}
Bond &$J$ & $K$ &$\Gamma$ &$\Gamma^{\prime}$ &$\xi$ &$\zeta$ \\
\hline
Z & -16.1 & 34.1 & 1.6 & -0.1 & 0.0 & 0.0  \\
X, Y & 9.4 & -18.9 & -0.7 & 2.2 & 1.1 & -4.7
\end{tabular}
\end{ruledtabular}
\end{table}

\section{Summary}

In summary, we have shown through a combination of symmetry analysis, structural relaxations and electronic structure calculations, that the experimentally observed insulating behavior in {\CIO} can originate in the charge ordering of iridium ions. In this scenario, nonmagnetic Ir$^{3+}$ and magnetic Ir$^{4+}$ ions alternate on the iridium honeycomb lattice. Similar charge order effects have been observed in 
K$_{0.5}$RuCl$_3$~\cite{Koitzsch2017}, with magnetic Ru$^{3+}$ and nonmagnetic Ru$^{2+}$ populating the transition metal honeycomb. Distinct from the K$_{0.5}$RuCl$_3$ case, we show that in Cu$_2$IrO$_3$ magnetic Cu$^{2+}$ ions are located in the iridium honeycomb voids, contributing to a novel, composite Cu$^{2+}$-Ir$^{4+}$- Kitaev honeycomb lattice that includes 3$d$ $s$=1/2 and 5$d$ $j_\mathrm{eff}$=1/2 magnetic moments. We note that the associated lattice distortions are small, and may be easily missed in diffraction experiments. Both our DFT and model spin Hamiltonian calculations support an antiferromagnetic ground state that has not been observed. However, the total energy gained by charge order is relatively small, suggesting that long-range charge order, and the respective magnetic order, is vulnerable to perturbation.

%\newpage
\acknowledgments
Y.L. acknowledges support from the National Natural Science Foundation of China (Grant No.~12004296). R.V. acknowledges support from the Deutsche Forschungsgemeinschaft (DFG, German Research Foundation) for funding through Project No. TRR 288 --- 422213477 (projects A05, B05). R.C. acknowledges support from the European Research Council under the European Union’s Horizon's 2020 Research and Innovation Programme Grant Agreement Number 788814 (EQFT). 
\bibliography{ref}

\appendix
\section{Structural parameters of \CIO}\label{app}
In this appendix we provide full crystallographic details for the $C2$ charge ordered and $C2/m$ reference crystal structures discussed in the main text. 

\begin{table*}[htpb]
\caption{Crystallographic details for the $C2$ charge ordered structure (left) derived from the $C2/m$ reference structure (right) refined by \emph{ab-initio} structural relaxation. The subgroup basis is \{[1,0,0],[0,1,0],[0,0,1]\}, origin=[0,0,0] with respect to the reference structure. Rows are spaced to indicate the relationship between sites of both structures (\emph{e.g.} site Cu2-2 of $C2/m$ splits into sites Cu2-2 and Cu2-3 of $C2$, while site Cu2-1 does not split). The lattice parameters are common to both structures.} 
\centering\def\arraystretch{1.1}
\label{c2}
\begin{ruledtabular}
\begin{tabular}{lcccc|lcccc}
\multicolumn{5}{c}{Space group $C2$ (No. 5)}& \multicolumn{5}{c}{Space group $C2/m$ (No. 12)}\\
\multicolumn{10}{c}{$a = 5.393~\mathrm{\AA}$, $b =  9.311~\mathrm{\AA}$, $c = 5.961~\mathrm{\AA}$, $\beta = 107.506^\circ$}\\
\hline
Atom & Wyckoff & $x$ & $y$ & $z$ & Atom & Wyckoff & $x$ & $y$ & $z$\\   
\hline                                                                             
Ir1 & 2a & 0 & 0.33386  & 0 & Ir & 4g & 0 & 0.33328 & 0 \\
Ir2 & 2a & 0 & 0.66615 & 0 & & & & \\[0.5cm]
Cu1 & 2a & 0 & 0.00000 & 0 & Cu1 & 2a & 0 &  0 & 0 \\[0.5cm]
Cu2-1 & 2b & 0 & 0.49718 & 0.5 & Cu2-1 & 2d & 0 &  0.5  & 0.5 \\[0.5cm]
Cu2-2 & 2b & 0 & 0.18094  & 0.5 & Cu2-2 & 4h & 0 &  0.17707  & 0.5 \\
Cu2-3 & 2b & 0 & 0.82302 & 0.5 & & & & \\[0.5cm]
O1  & 4c & 0.11950 &  0.49773 &  0.82439 & O1 & 4i & 0.10562 &  0.5 & 0.82306 \\[0.5cm]
O2  & 4c & 0.10630  & 0.17343 &  0.81942 & O2 & 8j & 0.11005 &  0.16919 & 0.82254 \\
O3  & 4c & 0.38529 &  0.32896  & 0.18061 & & & &\\
\end{tabular}                       
\end{ruledtabular}                                                                              
\end{table*}

\end{document}